\documentclass[12pt,twoside]{article}
\usepackage{fleqn,espcrc1}
\usepackage{graphicx}

\newcommand{\rts}{ \sqrt s}

\newcommand{\AmS}{{\protect\the\textfont2
  A\kern-.1667em\lower.5ex\hbox{M}\kern-.125emS}}

\title{$\eta$ nucleus optical potential in a chiral unitary approach }

\author{
T. Inoue\address{Institute f\"ur Theoretische Physik, Universit\"at T\"ubingen,
                 Auf der Mogenstelle 14, D-72076 Tuebingen, Germany}
and 
E. Oset\address{Departamento de F\'{\i}sica Te\'orica and IFIC, 
                Centro Mixto Universidad de Valencia - CSIC,
                Institutos de Investigaci\'on de Paterna,
                Apdo. 22085, 46071, Valencia, Spain}
       }
       
\begin{document}

\maketitle

\begin{abstract}
 The self-energy of an $\eta$ in a nuclear medium is calculated
in a chiral unitary model, and applied to $\eta$ states in nuclei.
Our calculation predicts an attractive
$\eta$ nucleus optical potential which can accommodate 
 many $\eta$ bound states in different nuclei.
 
\end{abstract}

\section{Introduction}

In the chiral unitary approach the 
S-wave scattering matrix is given by
\begin{equation}
   T(\rts)  = [1 - V(\rts)G(\rts)]^{-1} V(\rts)
\label{eqn:cum}
\end{equation}
in terms of the kernel matrix $V$ and the loop matrix $G$,
where $\rts$ is the invariant energy.
This is a solution of the Bethe-Salpeter equation which uses the fact that the 
off-shell part of the amplitudes can be absorbed in a renormalization 
of the couplings.
The kernel $V$ is given by the lowest order chiral Lagrangian,
while the loop $G$ is determined analytically by the unitarity condition with a
subtraction constant fitted to data.
We can obtain, for example, the $\eta n$ scattering amplitude from eq.(\ref{eqn:cum})
by considering the coupled channels: 
\{$\pi^- p$, \  $\pi^0 n$, \  $\eta n$, \  $K^0 \Lambda$, \ $K^+ \Sigma^-$, \ 
$K^0 \Sigma^0$, \ $\pi^0 \pi^- p$, and $\pi^+ \pi^- n$\}.

Analogously, the model gives scattering amplitudes in the nuclear medium by 
\begin{equation}
  T(P^0, \vec P ~; \rho) =  
  \left[ 1 - V(\rts) G(P^0, \vec P ~; \rho) \right]^{-1}  V(\rts)
\end{equation}
where $(P^0, \vec P)$ is a 4-momentum of the system and $\rho$ is the density of the medium. 
Here, the kernel is the same as in the free case, while the loop is replaced 
by the in-medium one,
which describes the propagation of systems in the medium
and hence includes medium effects.
In this case, the scattering matrix $T$ is no longer Lorentz invariant, 
and we need to fix a frame.

Supposing we have the in-medium $\eta n$ scattering amplitude,
as a function of 4-momentum in the nuclear matter rest frame,
we can obtain the self-energy of the $\eta$ theoretically
by summing the amplitudes over the nucleons in the Fermi sea
\begin{equation}
   \Pi_{\eta}(k^0, \vec k~; \rho) 
   =   
   4 \int^{\,k_F} \! \!  \frac{d^3 \vec p_n }{(2\pi)^3}~ 
   T_{\eta n}(P^0, \vec P ~; \rho) 
\end{equation}
where spin and isospin symmetry is assumed.
This self-energy determines $\eta$ states in nuclei,
which is investigated intensively nowadays
since it may provides new information about medium effects in hadrons.
In this paper, we calculate the $\eta$ self-energy in this way,
and comment about $\eta$ bound states in nuclei.  


\section{Our model}

In the meson-baryon 2-body sector our kernel matrix $V$ is given by 
\begin{equation}
 V(\rts)_{ij} = - C_{ij}\frac{1}{4 f_i f_j} (2 \rts - M_i - M_j)
            \sqrt{ \frac{M_i + E_i }{2 M_i} }
            \sqrt{ \frac{M_j + E_j }{2 M_j} }
\label{eqn:pot}
\end{equation}
which is derived from the lowest order chiral Lagrangian.
The coefficients $C_{ij}$ reflect $SU(3)$ flavor symmetry and 
are slightly modified by a vector meson exchange form factor. 
We use phenomenological values: 
$f_{\pi}=93$ MeV, $f_K = 1.22 f_{\pi}$ and $f_{\eta}=1.3 f_{\pi}$.  
For the S-wave $\pi N \leftrightarrow \pi \pi N$ sector,
we use the kernels which are obtained in our previous work,
where the inelasticities of $\pi N$ scattering are well reproduced
both in the isospin 1/2 and 3/2 channels \cite{Inoue}.

The in-medium loop functions are calculated by 
\begin{eqnarray}
 & & G_l(P^0,\vec P~;\rho) 
 \nonumber
 \\
 &=&
 i \!  
 \int \! \frac{d^4 q}{(2 \pi)^4} 
 \frac{M_l}{E_l(\vec P-\vec q)} 
 \frac{ \left(~ \theta(|\vec P-\vec q| - k_F ) ~\right) } 
      {P^0 - q^0 - E_l(\vec P-\vec q) - \left(~ \delta B(\rho) ~\right) } 
 \int_0^{\infty} \! \! \! d \omega
 \frac{2 \omega S_l(\omega,\vec q~;\rho)}{(q^0)^2-\omega^2}
 \\
 &\mbox{or}& 
 \nonumber
 \\
 &=&
 i^2 \! 
 \int \! \frac{d^4 q_1}{(2 \pi)^4} \! \!
 \int \! \frac{d^4 q_2}{(2 \pi)^4} 
 \frac{( \vec q_1 - \vec q_2 )^2\,M_N}{E_N(\vec P - \vec q_1-\vec q_2)}
 \frac{\theta( |\vec P-\vec q_1-\vec q_2|-k_f) }
       {P^0 - q_1^0 - q_2^0 -E_N(\vec P-\vec q_1-\vec q_2)}
 \nonumber
 \\
 & &
 \hspace{47mm} \times
 \int_0^{\infty} \! \! \! \! d \omega_1
 \frac{2 \omega_1 S_{\pi}(\omega_1,\vec q_1~;\rho)}{(q_1^0)^2 - \omega_{1}^2} \!
 \int_0^{\infty} \! \! \! \! d \omega_2
 \frac{2 \omega_2 S_{\pi}(\omega_2,\vec q_2~;\rho)}{(q_2^0)^2 - \omega_{2}^2}
\end{eqnarray}  
where the first line is for meson-baryon 2-body channels,
while the second line is for $\pi \pi N$ 3-body channels.
The $\theta$-function provides the Pauli blocking for nucleons.
The $\delta B(\rho)$ is the difference between the hyperon binding energy and
the nucleon one.
We use the conventional value: $+40 \, \rho/\rho_0$ MeV.
The part of the $\omega$-integral is the meson in-medium propagator in the Lehman representation,
where $S_l(\omega, \vec q~;\rho)$ is the spectral density of the meson, which is given by 
\begin{equation}
S_l(\omega, \vec q~;\rho) =  -\frac{1}{\pi} 
    \frac{\mbox{ Im}[ \Pi_l( \omega,\vec q~;\rho) ]}
         { | \omega^2 - \vec q\,^2 - m_l^2 
                 - \Pi_l(\omega,\vec q~;\rho) |^2 }
\end{equation}
with $\Pi_l( \omega,\vec q~;\rho)$ its self-energy.
We construct the pion self-energy from the $N$-hole, $\Delta$-hole and Roper-hole excitations,
where a recoil, a short range correlation and a form factor are introduced
in a traditional way. We use a theoretical 
$\Pi_K(\omega,\vec q~;\rho) = 0.13 m_K^2 \, \rho/\rho_0$ for the kaon \cite{OR}.
The self-energy of the $\eta$ which we calculate is used again in the calculation
iteratively untill convergence is reached, hence implementing selfconsistency in
the calculation.

The above loop integrals diverge and
we renormalize our model by means of cut-off regularization and subtractions.
We fix them so that in-medium loops coincide with free ones in the zero density limit.
The free loops can be fixed using scattering data.

\section{Results}

\begin{figure}[t]
\centering{ \includegraphics[height=101mm]{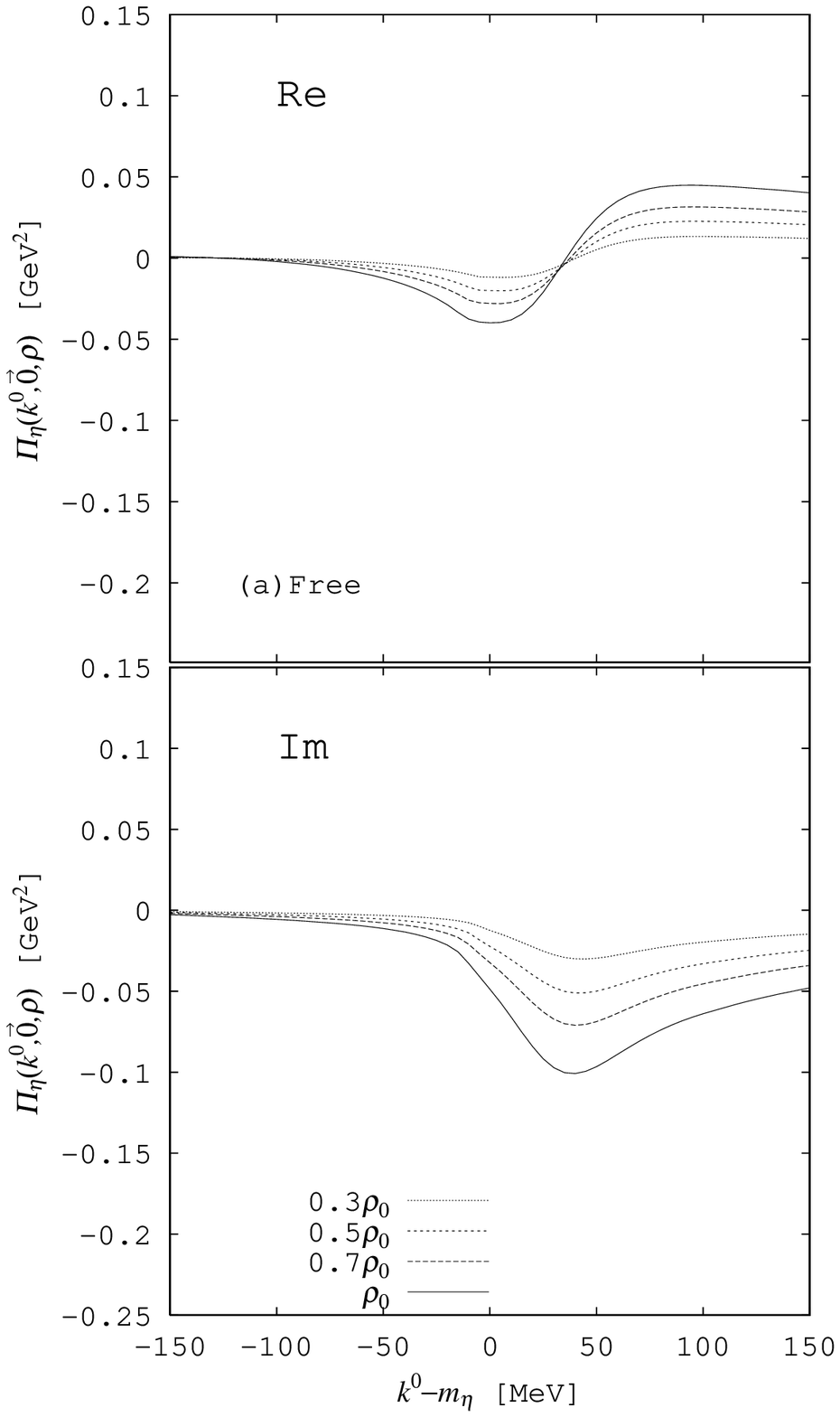}
            \includegraphics[height=101mm]{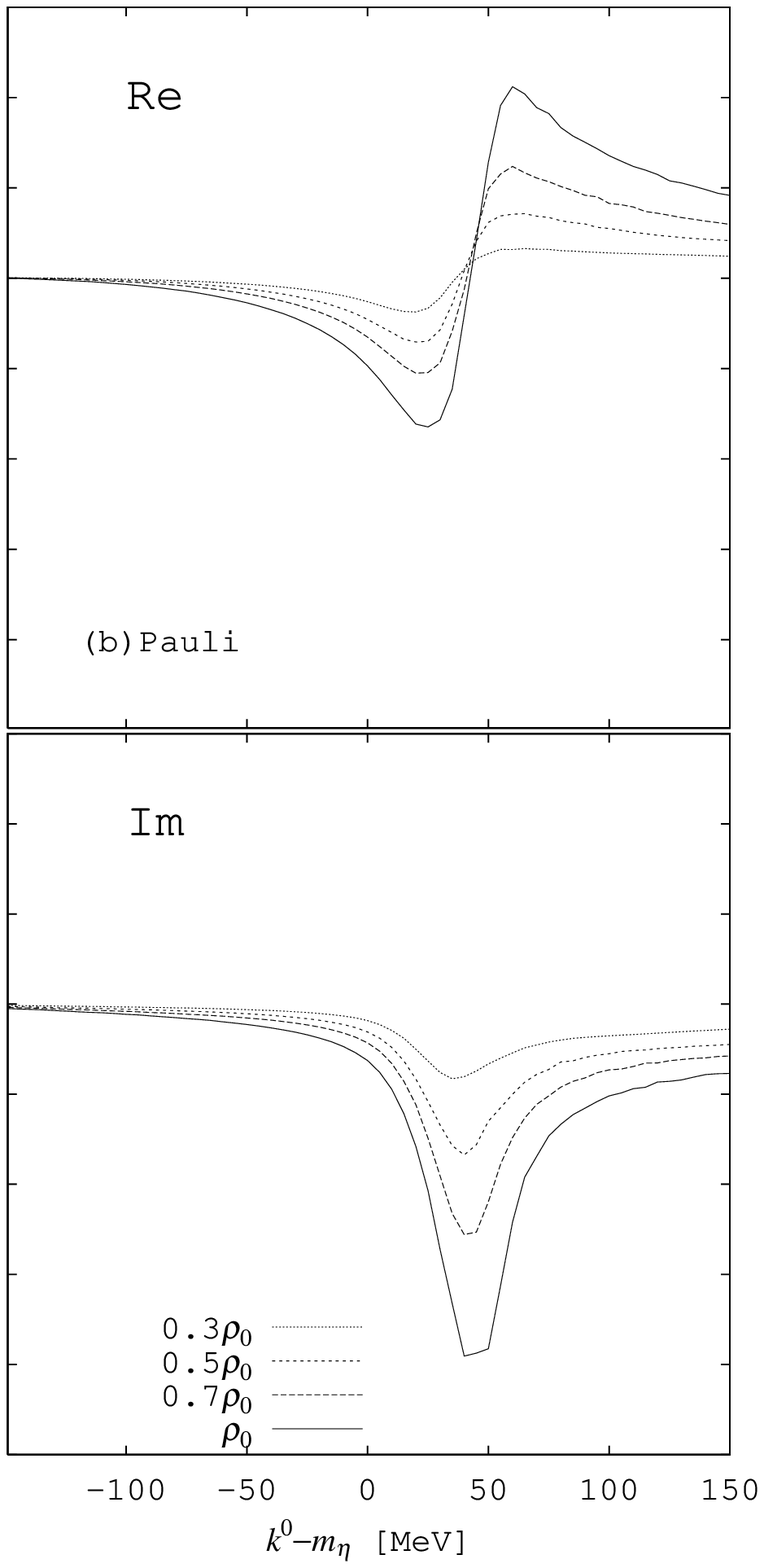}
            \includegraphics[height=101mm]{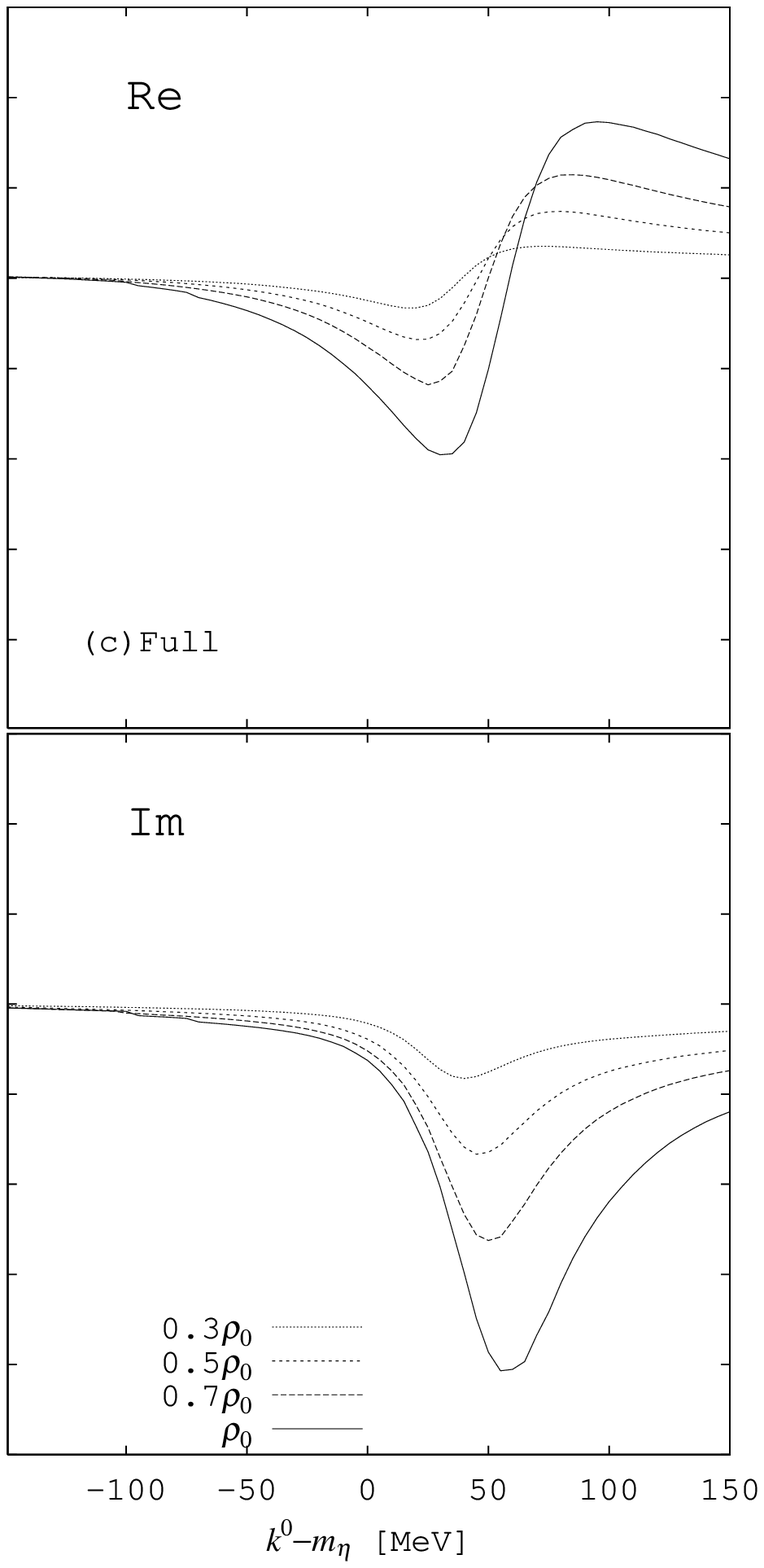}}
\caption{ Self-energy of $\eta$ with zero momentum as a function of energy,
  for four different densities and in three different approximations.
  In the left (a), the free space $\eta n$ amplitude is used.
  In the center (b), the Pauli blocking is taken into account.
  In the right (c),  both the Pauli blocking and hadron dressing are taken into account.}
\label{fig:iiabc}
\end{figure}

\begin{figure}[t]
%
\begin{minipage}[c]{90mm}
\includegraphics[width=75mm]{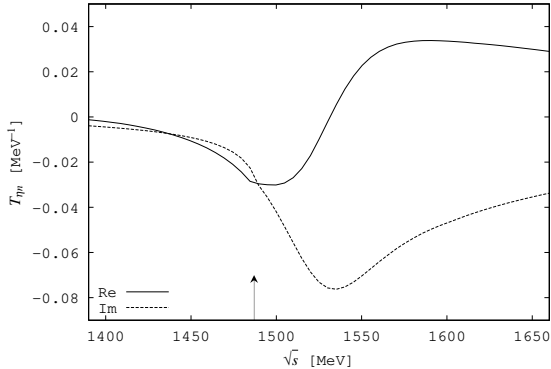}
\end{minipage}
\begin{minipage}[b]{60mm}
\caption{ Free space $\eta n$ scattering amplitude as a function of
           the invariant energy. The arrow shows the threshold.}
\label{fig:t66fr}
\end{minipage}
\end{figure}

Fig.\ref{fig:iiabc} shows the obtained self-energy of the $\eta$ with zero momentum 
as a function of energy.
We have calculated it with three different approximations for the nuclear medium.
The left (a) is obtained with the free space $\eta n$ scattering amplitude
shown in Fig.\ref{fig:t66fr}, 
and corresponds to so called $T\rho$ approximation with Fermi average.
The amplitude clearly shows the coupling to a resonance, $N^*(1535)$, 
which is generated in the present model.
Although we cannot confirm this amplitude by direct comparison with data,
it should be realistic because data of isospin 1/2 $\pi N$ scattering
are reproduced fairly well in the present model 
even in the $\eta n$ threshold region \cite{Inoue},
including a $N^*(1535)$ width of 93 MeV, which agrees with
the newest experimental data of $95 \pm 15$ MeV
at Beijing Electron-Positron Collider \cite{Bai}.
The $\eta n$ scattering length is $a_{\eta n}= 0.264 + i 0.245$ fm in this model. 

The center (b) is calculated with the Pauli blocking effects.
We see that the resonant shape is strongly enhanced compared to the left one (a).
This enhancement corresponds to a reduction of the resonance width
due to the Pauli blocking. On the other hand, 
we see that the position of the resonance,
or equivalently the mass of the resonance, is almost not shifted .
The reason is that the kaon-hyperon components dominate the resonance 
in the present model. This feature is in contrast to the $\Lambda(1405)$ case. 

The right (c) is our full calculation, where hadron dressing effects
are taken into account in addition to the Pauli blocking.
We see that the shape is broadened significantly.
This broadening is due to the distributed meson mass including the $\eta$ itself.
As a consequence, our final self-energy is quite different from the one in (a),
where a naive $T\rho$ approximation is used.

The following potential at the normal nuclear matter density
\begin{equation}
 U_{\eta}(\rho_0) \equiv \frac{\Pi_{\eta}(m_{\eta},\vec 0~;\rho_0)}{2 m_{\eta}}
 = -54 - i 29 ~\mbox{[MeV]}
\end{equation}
is helpful to get a feeling of our self-energy.
This is more attractive than other ones in the literature,
for example, $-20 - i 22$ MeV in ref. \cite{Wass}.

It is interesting to apply our self-energy to $\eta$ states in nuclei.
This has been done in \cite{Garcia-Recio:2002cu} by solving the Klein-Gordon equation using
the local density approximation to go from infinite matter to finite nuclei
\begin{equation}
  \left[ 
  -\vec\nabla^2 + \mu^2 + \Pi_{\eta}( \mbox{ Re}[E],\vec 0~; \rho(\vec r) )
  \right] \Psi 
  = E^2 \Psi
\end{equation}
where $\mu$ is the $\eta$ nucleus reduced mass. 
As shown in \cite{Garcia-Recio:2002cu}, many bound states are found in different 
nuclei,
where the  half widths of the bound states are comparable or even
larger than the separation between the levels.  According to these results, 
it is not easy to see these states in experiments unfortunately.
It was also shown in \cite{Garcia-Recio:2002cu} that
in heavier nuclei more several  bound states can appear
and that the optimal region to see bound states is around the $^{24}$Mg nucleus.

\end{document}